\documentclass[aps,pra,preprint,groupedaddress]{revtex4}
\usepackage{graphics}
\begin{document}
\title{Low noise amplification of an optically carried microwave signal:\\ application to atom interferometry}
\author{T.~L\'{e}v\`{e}que}
\author{A.~Gauguet\footnote{Present address: Department of Physics, Durham University, Rochester Building, South Road, Durham DH1 3LE, England}}
\author{W.~Chaibi\footnote{Present address: ARTEMIS - Observatoire de Nice, Boulevard de l'Observatoire, 06305 Nice, France}}
\author{A.~Landragin}

\email[]{arnaud.landragin@obspm.fr}
\affiliation{LNE-SYRTE, Observatoire de Paris, CNRS, UPMC, 61
avenue de l'Observatoire, 75014 Paris, FRANCE}

\date{\today}
%
%
%

%
%

%
\begin{abstract}
In this paper, we report a new scheme to amplify a microwave
signal carried on a laser light at $\lambda$~=~852~nm. The
amplification is done via a semiconductor tapered amplifier and
this scheme is used to drive stimulated Raman transitions in an
atom interferometer. Sideband generation in the amplifier, due to
self-phase and amplitude modulation, is investigated and
characterized. We also demonstrate that the amplifier does not
induce any significant phase-noise on the beating signal. Finally,
the degradation of the performances of the interferometer due to
the amplification process is shown to be negligible.
\end{abstract}

\maketitle

\section{Introduction}
\label{intro} Optically carried microwave signals are of special
interest in a large field of applications, from optical
communication to opto-electronics techniques for detection such as Lidar-Radar~\cite{morvan02}, and to fundamental metrology
and spectroscopy. For instance, time and frequency dissemination at long
distances by optical fibres has shown extremely high
performances~\cite{daussy05}. Furthermore, this principle is also widely used in atomic physics. As an example, it is at the basis of the
Coherent Population Trapping (CPT) clocks~\cite{CPT}, which
benefits from the reduction of size compared to standard microwave
clocks. This method is also used in most of the atom
interferometers to generate Raman lasers for manipulating atomic
wave-packets~\cite{kasevich91}. In fact, it enables reduction of the
propagation noise over the laser paths and systematic errors. The
signal is composed of two optical frequencies separated by a frequency
in the microwave range. This can be achieved by different means:
directly from a two frequency laser~\cite{morvan02}, by a single
sideband electro-optic modulator~\cite{seeds02}, by filtering two
sidebands from an optical comb~\cite{goldberg83,kefelian06} or
from a phase-lock between two independent
lasers~\cite{santarelli94}, as used in this work. In most of the
applications, these sources are not powerful enough and would
benefit from being amplified without adding extra sidebands or
extra phase noise onto the microwave signal.

\begin{figure}[!h]
\centering \resizebox{8.5cm}{!}{
\includegraphics{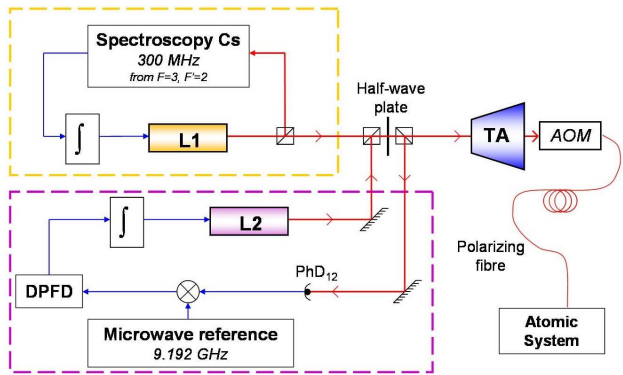}}
\caption{Laser setup. The frequency-reference laser L1 is locked
on a spectroscopy signal. The L2 laser frequency is mixed with the
frequency-reference laser and the optical beat note is compared to
a microwave reference and phase-locked through a Digital Phase and
Frequency Detector (DPFD). L1 and L2 are combined on a polarizing
beam splitter cube and amplified using the same tapered amplifier
(TA). The output power is injected in a polarizing fibre through
an acousto-optical modulator (AOM).} \label{bancraman}
\end{figure}

In this paper, we report a study of the influence of a
semi-conductor tapered amplifier on a two frequency laser system.
This setup is dedicated to the generation of a pair of Raman
lasers at $\lambda$~=~852~nm, with a fixed frequency difference
close to 9.192~GHz for further use in an atom
interferometer~\cite{canuel06}. The experimental setup is
described in section
II. Sections III and IV are dedicated to the characterization and measurement of the optical spectrum and to the analysis of the spurious sideband generation due to non-linear effects in the gain medium~\cite{ferrari99}. Then we measure the extra noise added by the amplifier on the microwave signal in section V. Finally, the impact on our atom interferometer is quantified in section VI.\\

\section{Experimental setup}

The laser setup consists of two external-cavity laser diodes using
SDL~5422 chip. These sources are based on an intracavity
wavelength selection by an interference-filter~\cite{baillard06}
and benefit from a narrow linewidth (14~kHz) and a wide tunability
(44~GHz). The diodes are regulated around room temperature and
supplied by a current of 80~mA to provide an optical output power
of 45~mW. The frequency locks are achieved with a feedback to the
diode current and the length of the external cavity.

The first laser L1 (Fig.~\ref{bancraman}), is used as an absolute
frequency reference. It is locked 300~MHz red detuned from the
atomic transition between the $|6S_{1/2},F=3\rangle$ and\\
$|6P_{3/2}, F=2\rangle$ states of Caesium (D2 line) using a
frequency modulation spectroscopy technique~\cite{hall81}.

The phase difference between L1 and L2 is locked with the
method described in~\cite{cheinet06} and summarized in the
following. Small amounts of light of L1 and L2 are superimposed on a fast
photoconductor (PhD$_{12}$, Hamamatsu G4176, bandwidth: 15~GHz).
The beat note at $\nu _{12}$~=~9.192~GHz is then mixed with a
reference signal, given by a microwave synthesizer~\cite{yver04}.
The output signal is sent to a Digital Phase and Frequency
Detector (DPFD, MCH~12140) which derives an error signal
proportional to the phase difference between the two lasers. After
shaping and filtering, this output signal is used to generate the
feedback allowing to phase-lock the laser L2 on L1. In this way, the
features of the microwave reference is mapped on the optical
signal with a bandwidth of 3.5~MHz.

In order to provide sufficient optical power, the output signals of L1 and
L2 are injected with the same linear polarization in a GaAs
tapered semiconductor waveguide amplifier (TA,
EYP-TPA 0850-01000-3006 CMT03) pumped by a current of $I$~=~2~A
and stabilized to room temperature. A half-wave plate and a
polarizing cube allow the power ratio between the two lasers to be
adjusted at the input of the TA. In a normal operation, 11.2~mW of
L1 and 16.2~mW of L2 are injected in the TA, which runs in a
saturated gain regime. Then the output beam (with a power of
770~mW) passes through an acousto-optical modulator (AOM),
driven at 80~MHz, from which the first output order is coupled
to a polarizing fibre. This scheme enables the laser light to be
pulsed on the atomic system by switching the RF signal of the AOM.

\section{Self-phase and amplitude modulation}

In this part, we study the sideband generation due to simultaneous
phase and amplitude modulation in the gain medium. As two closely
detuned optical frequencies are injected in the TA, it is crucial
to determine the spectral impact of potential non-linear effects
in the semiconductor during the amplification process. Indeed, the
beat note at $\nu_{12}=$ 9.192~GHz between L1 and L2 induces a
strong modulation of the power through the TA which affects the
gain and the optical index in the semiconductor~\cite{ferrari99}.
In this situation, the resulting sidebands could cause undesirable
effects on our experiment: for instance detuned excitations can
shift the energy levels of our atomic system, called light
shift~\cite{weiss94}, or it could drive unwanted Raman
transitions. For this reason, we conducted simulations which were
compared to experimental measurements.

The total electric field propagating through the TA can be
described by,

\begin{eqnarray}
\label{general} E\left(t,z\right) &=& A\left(t,z\right)
e^{-i\left(\omega_0 t-kz\right)},\\
\nonumber &=&
\left|A\left(t,z\right)\right|e^{i\psi\left(t,z\right)}\times
e^{-i\left(\omega_0 t-kz\right)},
\end{eqnarray}

where $A\left(t,z\right)$ is the wave envelope which varies at the
microwave frequency, $k$ is the wave vector, and $\omega_0$ is the
optical carrier frequency. By referring
$P\left(t,z\right)=\left|A\left(t,z\right)\right|^2$ to the
envelope power ($\psi\left(t,z\right)$ is its phase), we obtain at
the TA input the modulation profile,

\begin{equation}
\label{PIN} P\left(t,z=0\right)=P_{0}\left(1+m \cdot \cos
\left(2\pi \nu_{12} t+\phi\right) \right),
\end{equation}

where $P_{0}$ is the nominal modulation power, $m$ is the
modulation factor and $\phi$ is the phase difference between L1
and L2 (see Appendix \ref{equations_interactions}). In our case,
we have $P_0 \simeq 27.4 $~mW and $m \simeq 0.983$. The amplifier
is then driven between a saturated state and a non-saturated state
at the frequency $\nu_{12}$. In order to calculate the sideband
generation expected from the amplification process, we write, as
for any amplification medium, the interaction between the carriers
and the light field equations. This was widely described in
reference~\cite{agrawal89,omahony88} for the case of a constant
cross section amplifier. Taking into account the amplifier splay
(see appendix~\ref{equations_interactions}), these equations
become,

\begin{eqnarray}
\label{Equation_systeme_P}
\frac{\partial P}{\partial z}+\frac{1}{v_g} \frac{\partial P}{\partial t}=gP,\\
\label{Equation_systeme_phi}
\frac{\partial \psi}{\partial z}+\frac{1}{v_g} \frac{\partial \psi}{\partial t}=-\frac{1}{2}\alpha g,\\
\label{Equation_gain_g} \frac{\partial g}{\partial t}=
\frac{g_0-g}{\tau_c}-\frac{gP}{E\mathrm{_{sat0}}\left(1+\mu z\right)},
\end{eqnarray}

where $\tau_{C}$ is the carrier lifetime in the semiconductor,
$E\mathrm{_{sat0}}$ the saturation energy for the initial amplifier cross
section, $\mu$ is the amplifier splay factor, $\alpha$ the
linewidth enhancement factor and $v_g$ is the wave group velocity.
$g$ refers to the linear gain and the small-signal
gain $g_{0}$ is defined as,

\begin{equation}
\label{g0} g_{0}=\Gamma a N_{0} \left(\frac{I}{I_{0}}-1\right),
\end{equation}

where $I_{0}$ and $N_{0}$ are the current and carrier density
required for transparency, $a$ the gain coefficient and $\Gamma$
the confinement factor (see
Appendix~\ref{equations_interactions}).

The sideband generation is due to a phase and an amplitude
modulation resulting from a non-linear gain modulation. This gain
modulation depends on the field amplitude along the amplifier and
lead to a non-linear distortion of the signal at the output of the
TA. A modification of the optical
index~(\ref{Equation_systeme_phi}) induces an optical phase
modulation. Indeed, an increase of the gain modulation exacerbates
the phase and the power modulation distortions at the same time.

The evaluation of this effect requires the set of equations
(\ref{Equation_systeme_P}-\ref{Equation_gain_g}) to be solved.
Since the relaxation time $\tau_c$ and the excitation
characteristic time $1/\nu_{12}$ are of the same order of
magnitude, the usual adiabatic approximation can't be used.
Therefore, the system has been numerically solved to obtain the
steady state electric field at the output of the TA for comparison
with experimental measurements.

\section{Optical spectrum measurement}

Due to the weak power expected in each sideband, usual methods to
measure optical spectrum such as the transmission through a
Perot-Fabry cavity or the diffraction on a high resolution grating
are not suitable. Instead, we use the beat note with a close known
optical frequency in order to achieve the precise measurement of
the different components of the laser beam. The output signal of a
beat note contains a frequency component corresponding to the
difference between the two mixed frequencies. It can be measured
with a fast photodetector. The power of the beat note is then
proportional to the product of the fields of the two components.

\begin{figure}[!h]
\centering \resizebox{8.5cm}{!}{
\includegraphics{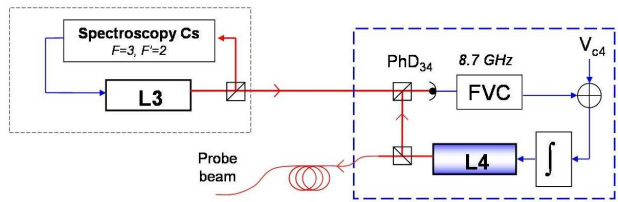}}
\caption{Probe laser setup. The frequency-reference laser L3 is
locked on a spectroscopy signal. The L4 laser beam is mixed with
the L3 beam and the beat note signal at 8.7~GHz is servo looped
through a Frequency to Voltage Converter (FVC). The L4 laser, used
to generate the probe beam, is injected in a fiber.}
\label{lasersonde}
\end{figure}

\begin{figure}[!h]
\centering \resizebox{7cm}{!}{
\includegraphics{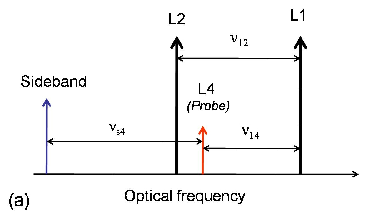}}
\resizebox{7cm}{!}{
\includegraphics{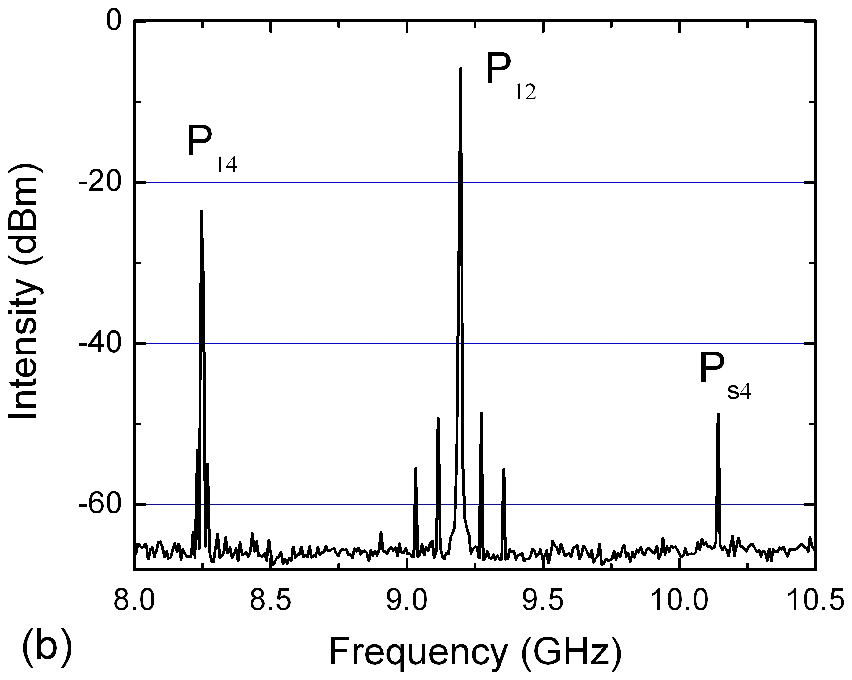}}
\caption{(a) Position of the different components in the optical
spectrum. (b) Measured beat note between L1, L2 and the probe
laser L4 at the output of the polarizing fibre. Three main
frequencies are obtained: $\nu_{12}$ (9.192~GHz), $\nu_{14}$
(8.245~GHz) and $\nu_{s4}$ (10.142~GHz). } \label{beatnote}
\end{figure}

In order to realize a spectrum measurement, we set up a probe
laser L4. As shown on Fig.~\ref{lasersonde}, the frequency of L4
can be adjusted around the Cesium D2 line by the comparison to a
reference laser L3. This frequency lock is achieved by measuring
the beat note of the two lasers, through a Frequency to Voltage
Converter (FVC). A reference DC voltage (VC4) allows the frequency
detuning between L3 and L4 to be set around 8.6~GHz. As a result,
L4 is detuned from L1 by $\nu_{14}$ = 8.3~GHz
(Fig.~\ref{beatnote}a). The optical frequency spectrum of the
phase-locked lasers (L1+L2) can be analyzed by superimposing the
probe beam L4 on the photoconductor PhD$_{12}$, which exhibits the
microwave components of the beat note between L1, L2 and L4. As
explained before, mixing three optical frequencies on a
photodetector gives rise to three frequencies in the microwave
domain. In this setup, L2 and L4 are sufficiently close that the
beat note between these two lasers~(950~MHz) is filtered out by
the detection bandwidth. Thus, we verify that before optical
amplification, the microwave spectrum of the PhD$_{12}$ signal is
composed of a component at $\nu_{12}$ of power $P_{12}$ and a
component at frequency $\nu_{14}$ of power $P_{14}$. In order to
investigate impact of the amplification on the spectral content,
the same beat note is measured again at the output of the
polarizing fibre and displayed on~Fig.~\ref{beatnote}b. Compared
to the expected microwave spectrum, containing the two previous
frequencies ($\nu_{12}$ and $\nu_{14}$), it exhibits an additional
signal at $\nu_{s4}$~=~10.142~GHz of power $P_{s4}$ which is
related to the beat note between L4 and an additional sideband
detuned by 9.197~GHz from L2. Another symmetric sideband at
9.192~GHz above L1 leads to a beating signal with L4 out of the
bandwidth of detection. Small spurious sidebands around $\nu_{12}$
are generated by the acousto-optic modulator~(AOM) at 80~MHz and
can be ignored in the following.

\begin{figure}[!h]
\centering \resizebox{8cm}{!}{
\includegraphics{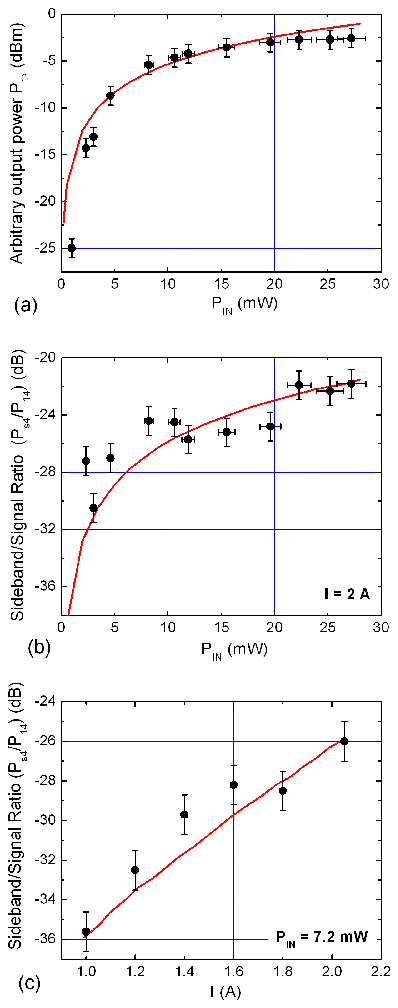}}
\caption{(a) Evolution of the micro-wave signal at frequency
$\nu_{12}$ versus the total optical input power. Variation of the
ratio between the microwave signals  $P_{14}$ and $P_{s4}$ at
frequencies $\nu_{14}$ and $\nu_{s4}$ (i.e. ratio of the electric
field in the sideband and in L1) as a function of the total
optical input power in the TA (b) and of the input current in the
TA (c).} \label{measure}
\end{figure}

The TA characterization requires the determination of the set of
parameters ($g_0, E\mathrm{_{sat0}},\alpha,I_0, a,N_0$). For this
purpose, we perform three measurements, which are displayed in
Fig.~\ref{measure}. First, we measure $P_{12}$ as a function of
the total optical input power~(Fig.~\ref{measure}a). Then we
record the two magnitudes $P_{14}$ and $P_{s4}$ as a function of
the total input optical power~(Fig.~\ref{measure}b) and the
current~(Fig.~\ref{measure}c). From $P_{s4}/P_{14}$, we deduce the
ratio between the fields of L1 and the sideband.

Using (Fig.~\ref{measure}a) and (Fig.~\ref{measure}b), we infer
$g_0\simeq 1.33\times 10^3$~m$^{-1}$, $E\mathrm{_{sat0}}\simeq
8$~pJ and $\alpha \simeq 6$, which are in agreement with the
values indicated in reference~\cite{agrawal89}. From
$E\mathrm{_{sat0}}$, we get $a = 5.2\times 10^{-15}$~cm$^2$. Then,
we use (Fig.~\ref{measure}c) to work out $I_0 \simeq 0.12$~A and
$N_0 \simeq 1.8\times 10^{20}$~m$^{-3}$. It gives the active
volume $V\simeq 8\times 10^{-13}$~m$^3$, in good agreement with
the geometrical calculation $V\simeq 5\times 10^{-13}$~m$^3$.

In normal operation, the ratio ($P_{s4}/P_{14}$) of the micro-wave
power is of -22~dB, corresponding to an optical power ratio
between the sideband and L1 of $4\times 10^{-5}$. As expected,
increasing the current or the input power leads to an increase in
sideband generation. Finally, we work out from the simulation that
the index modulation is responsible for 90 percent of the sideband
generation and the amplitude modulation is responsible for the
last 10 percent.
\section{Phase noise measurement}

The residual phase noise between the two lasers is crucial for the
sensitivity of our experiment. Indeed, the laser phase difference
is imprinted on the atom wave-function during the stimulated Raman
transitions~\cite{cheinet08}. The effective phase noise affecting
the interferometer is the sum of different contributions: the
noise induced by the microwave frequency reference~\cite{yver04},
the residual noise in the phase-lock loop~\cite{cheinet06} and the
noise added along the propagation through the TA and the fibre.
The present study is focused on the characterization of these two
last terms.

\begin{figure}[!h]
\centering \resizebox{8cm}{!}{
\includegraphics{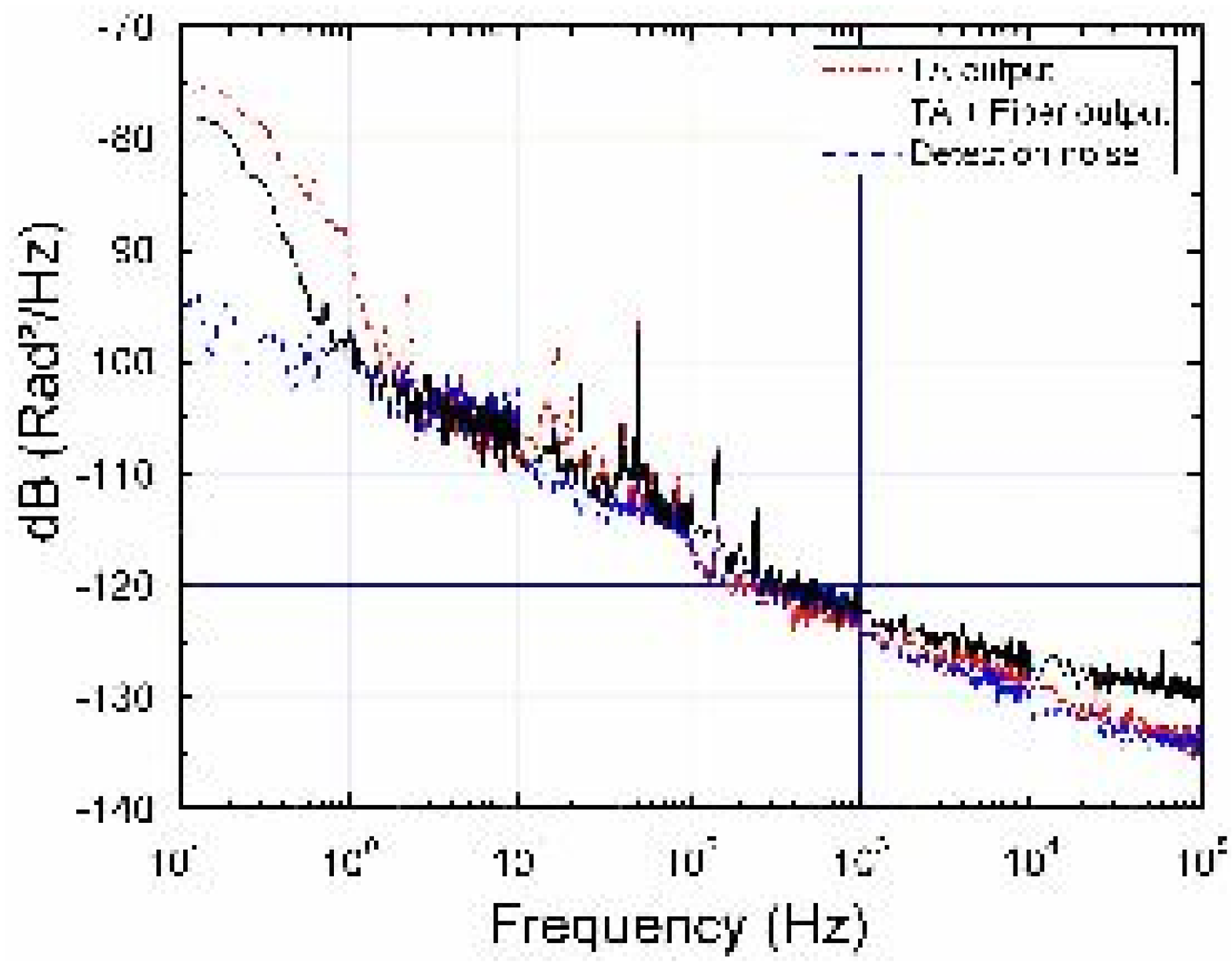}}
\caption{Phase noise power spectral density. The dash blue curve
shows the noise from the detection system and the dotted red and
black full curves respectively represent the noise after the TA
and the fiber.} \label{phasenoise}
\end{figure}

The additional phase noise is measured by mixing the beat notes of
the two lasers (L1 and L2) before and after the propagation along
the TA and the fibre. The two microwave signals are then amplified
and combined together on a mixer~(ZMX-10G). Their relative phase
$\phi_{A}$ is adjusted with a phase-shifter and set around $\pi/2$
to reach the optimal phase sensitivity at the mixer output. After
filtering the high frequency components, the output signal is
given by,

\begin{equation}
\label{smixer} s_{Mixer}=K_{d} \cos \left(\tilde{\phi_{n}} +
\phi_{A} \right) \approx K_{d} \cdot \tilde{\phi_{n}},
\end{equation}

where $K_{d}$ represents the scaling factor (0.3~V/rad) between
the phase shift and the output level of the mixer.
$\tilde{\phi_{n}}$ is the phase noise between the two beat notes.
The measurement of the power spectral density of phase noise
$S_\phi$ is obtained with a FFT analyzer and displayed on
Fig.~\ref{phasenoise}. It exhibits the phase noise contribution
induced by the TA and the fibre. The measurement is compared to
the detection noise, which is obtained using the same signal in
both inputs of the mixer. This detection noise represents the
lower limit of the phase noise which can be detected by this
setup. No significant additional phase noise from the TA and the
fibre is measured above 1~Hz.

\section{Impact on an atom interferometer}

\begin{figure}[!h]
\centering \resizebox{8cm}{!}{
\includegraphics{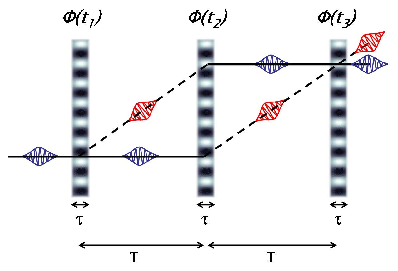}}
\caption{Scheme of the $\pi/2$-$\pi$-$\pi/2$ interferometer. The
Raman beam splitters of duration $\tau$ are separated by a free
evolution time~$T$. Atomic wave-packets are split, deflected and
recombined by the Raman lasers to realize the equivalent of a
Mach-Zehnder interferometer.} \label{interferometer}
\end{figure}

Raman transitions enable the manipulation of atomic wave-packets
by coupling two internal states of an atom using a two photon
transition~\cite{kasevich91}. In order to drive these transitions,
two lasers with a difference of frequency in the microwave range
are needed. This laser setup is implemented to realize the
functions of mirrors and beam splitters for atoms~\cite{borde91}
in a six-axis inertial sensor~\cite{canuel06}. A sequence of three
optical pulses ($\pi/2$, $\pi$, $\pi/2$) of duration $\tau$,
separated by free propagation time~$T$ is used to create the
equivalent of a Mach-Zehnder
interferometer~(Fig.~\ref{interferometer}) for
atoms~\cite{borde02}. Here, we estimate the possible limitations
induced by the common amplification of the two Raman lasers in our
specific device which come from unwanted Raman transitions due to
the sideband generation or additional phase noise in the TA.

First, the sideband generation in the TA is small enough to not
give rise to significant Raman transitions between laser L1 or L2
and one of the sidebands. The diffraction process is characterized
by the effective Rabi frequency~\cite{moler1992} which scales as
the product of the fields of the two lasers and the inverse of the
Raman detuning. The amplitude of the sideband is 22~dB smaller and
9~GHz further detuned than the main optical fields. Therefore the
corresponding Rabi frequency is reduced by almost four orders of
magnitude.

Second, we estimate the additional phase shift from the TA. The
phase shift at the output of the interferometer is deduced from
the transition probability of the atoms between the two coupled
states. It depends on the inertial forces and on the the phase
noise between the two lasers~$\Delta\phi\mathrm{_{laser}}$
~\cite{cheinet08}, as the phase difference between L1 and
L2~($\phi(t_i)$) is imprinted on the atomic wave function at the
moment of each pulse,

\begin{equation}
\label{H} \Delta\phi\mathrm{_{laser}}=\phi(t_1)-2\phi(t_2)+\phi(t_3).
\end{equation}

The transfer function of the spectral power density phase noise
simplifies at low frequency ($f \ll 1/\tau$) as,

\begin{equation}
\label{H} \mid H(2\pi f)\mid ^{2} = \frac{4}{\pi^{2} \cdot f^{2}}
\sin^{4} \left(\pi f T \right),
\end{equation}

where $T$, the interaction time, is typically 40~ms. For high
frequencies, the transfer function decreases as a second order low
pass filter with effective cutoff frequency ($1/(4\surd3\tau)$).
The total phase noise added by the TA and the fibre is given by
the integration over the full spectral range of $S_\phi$ weighted
by the transfer function. It leads to fluctuations on the output
atomic phase noise of~0.21~mrad from shot to shot. This
contribution is negligible compared to the noise generated by the
present microwave synthesizer (2.51~mrad), or best frequency
synthesizer based on current quartz oscillator ($\sim$~1~mrad).

\section{Conclusion}

To conclude, we demonstrate that amplifying optically carried
microwave signal with a tapered amplifier induces neither
microwave phase noise nor significant spurious optical sidebands.
This setup does not give any additional limitation on the
sensitivity of our device, which uses relevant parameters for cold
atom interferometer. The sideband generation has been carefully
characterized and shows a good agreement with the calculations.
This effect becomes more significant as the microwave frequency
get smaller. For instance, the model shows that similar results
should be found for an interferometer based on Rb atoms (microwave
frequency of 6.834~GHz), with an increase of the sideband
generation as small as 2~dB.

To a great extent, it shows that a powerful optically carried
microwave signal can be realized in two stages: one performing the
optical carried microwave signal and the other supplying the
optical power. The quality of the amplification does not depend on
a particular method to generate the optical signal and should be
similar when sideband generation in an electro-optic modulator is
used to generate the microwave signal. This can lead to
simplifications of optical systems in atom
interferometry~\cite{nyman06} and of amplification for frequency
and time dissemination at long distance. In addition, this method
could simplify atom interferometer experiments in which several
separated powerful Raman beams are required~\cite{muller08}, as
they can be amplified independently without special care. Finally,
this study has been realized in the steady state regime, and might
be extended to the pulsed regime~\cite{Takase2007} which allows for
higher laser power.

\begin{acknowledgments} We would like to thank G. Lucas-Leclin for
fruitful discussions, E. Foussier for her contribution to the
experimental setup, J. D. Pritchard and C. Garrido Alzar for
careful readings. We also thank the Institut Francilien pour la
Recherche sur les Atomes Froids (IFRAF) and the European Union
(FINAQS STREP NEST project contract no 012986 and EuroQUASAR IQS
project) for financial support. T.L. thanks the DGA for supporting
his work. W.C. thanks IFRAF for supporting his work.
\end{acknowledgments}


\appendix

\section{Light-semiconductor interaction equations}
\label{equations_interactions}

\begin{figure}[!h]
\centering
\resizebox{6cm}{!}{\rotatebox{-90}{\includegraphics*[6mm,18mm][196mm,278mm]{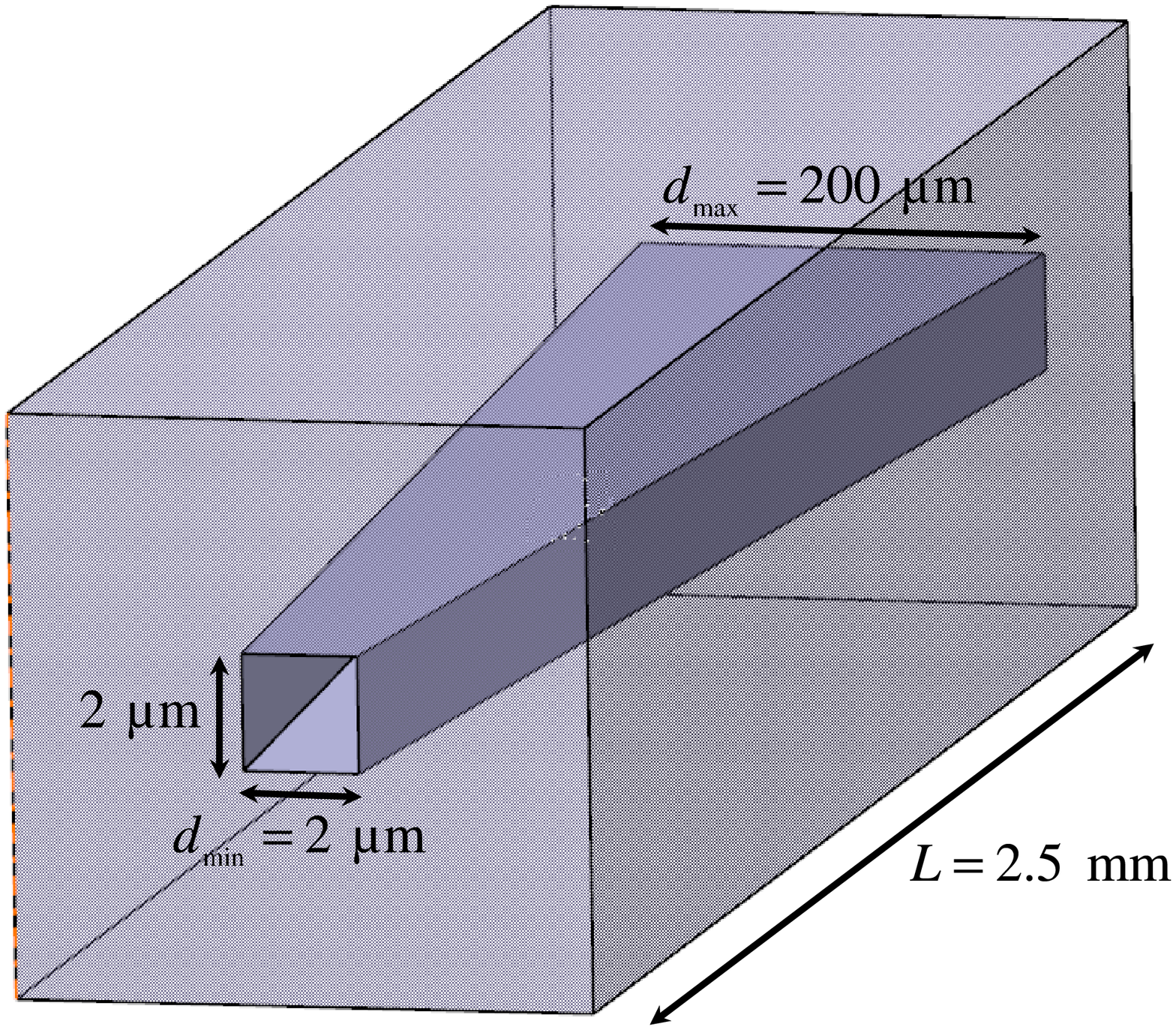}}}
\caption{Scheme of the tapered amplifier (TA). $d_{\mathrm{min}}$
and $d_{\mathrm{max}}$ are the input and the output horizontal
width of the waveguide.} \label{Amplificateur}
\end{figure}

The theory of self phase modulation in semiconductor amplifiers
has been developed for pulse propagation in
reference~\cite{agrawal89} to the case where the pulse width
$\tau_p$ is assumed to be much larger than the interband
relaxation time $\tau_{in} = 0.1$~ps which governs the dynamics of
the induced polarization. Our case concerns the beat note of two
lasers, whose frequencies are separated by the beat frequency
$\nu_{12}$, and amplified through a tapered semiconductor
$L=2.5$~mm long (\ref{Amplificateur}). In
reference~\cite{agrawal89}, it is shown that the dynamic of a
semi-conductor with a constant cross section is given by the set
of equations,

\begin{eqnarray}
\label{Equation_systeme_P2}
\frac{\partial P}{\partial z}+\frac{1}{v_g} \frac{\partial P}{\partial t}=gP,\\
\label{Equation_systeme_phi2}
\frac{\partial \psi}{\partial z}+\frac{1}{v_g} \frac{\partial \psi}{\partial t}=-\frac{1}{2}\alpha g,\\
\label{Equation_gain_g2} \frac{\partial g}{\partial t}=
\frac{g_0-g}{\tau_c}-\frac{gP}{E\mathrm{_{sat0}}},
\end{eqnarray}

In our case, the electric field input is given by,

\begin{eqnarray}
E\left(t,z = 0\right) &=& E_1 e^{-i(\omega_1 t-\phi_1)}+E_2
e^{-i(\omega_2 t-\phi_2)},\\
\nonumber &=& A\left(t,0\right)\times e^{-i\omega_0 t},
\end{eqnarray}

where the envelope $A\left(t,0\right)$ at the amplifier input is
given by,

\begin{equation}
\label{Amplitude} A\left(t,0\right) =
\left[P_0\left(1+m\cos\left(2\pi\nu_{12}t+\phi\right)\right)\right]^\frac{1}{2}\times
e^{i\psi(t,0)},
\end{equation}

\begin{eqnarray}\nonumber
\psi\left(t,0\right) &= \phi_0+\arctan \left(\sqrt{\frac{1-m}{1+m}}\tan\left(\pi\nu_{12}t+\frac{\phi}{2}\right)\right)\\
 &+\pi \times
\textrm{floor}\left(\frac{\pi\nu_{12}t+\frac{\phi}{2}}{\pi}\right),
\end{eqnarray}

where $P_0 = E^{2}_{1}+E^{2}_{2}$, $m =
\frac{2E_1E_2}{E^{2}_{1}+E^{2}_{2}}$, and $\phi =
\left(\phi_1-\phi_2\right)$, and where we set $\omega_0$ equal to
$\left(\omega_1+\omega_2\right)/2$ and $\phi_0$ to
$\left(\phi_1+\phi_2\right)/2$.

These equations are inferred from a combination of the
electromagnetic field propagation equation, on which the slowly
varying envelope approximation has been applied, and the carrier
density equation. In the latter, the carrier density diffusion has
been neglected in the $z$ direction (the carrier diffusion length
$d\mathrm{_{diff}}\simeq 40\mu$m$<<L$). However, it tends to make
uniform the carrier density in the transverse directions
($d\mathrm{_{diff}}>>d_{\mathrm{min}}$). The carrier density
equation is therefore averaged over the directions $x$ and $y$ and
appears through the factor,

\begin{equation}
\label{Gamma} \Gamma =
\frac{\int^{d_{\mathrm{min}}}_{0}\int^{d_{\mathrm{min}}}_{0}{\left|F(x,y)\right|^2
dx dy}} {\int^{\infty}_{-\infty}
\int^{\infty}_{-\infty}\left|F(x,y)\right|^2 dx dy},
\end{equation}

where $F(x,y)$ represents the transverse mode.

In the case of a TA, as the amplifier get larger, the carrier
diffusion tends to make the carrier density uniform, which
distorts the transverse mode. Nevertheless, since the mode is kept
guided, the factor $\Gamma$ remains of the order of 0.9 and is
considered constant in the following. Therefore, the set of
equation [\ref{Equation_systeme_P2}-\ref{Equation_gain_g2}] is
still valid in the case of a tapered amplifier provided we change
the last equation to,

\begin{equation}
\label{g_evase} \frac{\partial g}{\partial t}=
\frac{g_0-g}{\tau_c}-\frac{gP}{E\mathrm{_{sat0}}\left(1+\mu z\right)},
\end{equation}

where $\mu = \left(d_{\mathrm{max}}-d_{\mathrm{min}}\right)/L$
describes the amplifier splay. It has to be noticed that all
losses and dispersion effects have been neglected.

In order to solve the set of equations
(\ref{Equation_systeme_P2},\ref{Equation_systeme_phi2},\ref{g_evase}),
we consider the amplifier as a succession of equally $\Delta z$
long subdivisions of fixed transverse cross section. The
combination of equation (\ref{g_evase}) and equation
(\ref{Equation_systeme_P2}) can be analytically integrated over
$z$ along each subdivision and then numerically solved according
to the shifted time $t-z/v_g$. Afterwards, equation
(\ref{Equation_systeme_phi2}) is easily integrated to obtain the
electric field $E\left(z,t\right)$. Using a Fourier transform, we
compute the electric field spectrum at the TA output.

%
%

\end{document}